\documentclass[%
 reprint,
 amsmath,amssymb,
 aps,
]{revtex4-2}

\usepackage[dvipsnames, svgnames, x11names]{xcolor}
\usepackage{graphicx}% Include figure files
\usepackage{dcolumn}% Align table columns on decimal point
\usepackage{bm}% bold math
\usepackage{amsmath}
\usepackage{multirow}
\usepackage{booktabs}
\usepackage{amsfonts}
\usepackage{makecell}
\usepackage{float}
\usepackage{diagbox}
\usepackage{tabularx}
\begin{document}

\preprint{APS/123-QED}

\title{Proposal for practical Rydberg quantum gates using a native two-photon excitation }% Force line breaks with \\

\author{Rui Li$^{1,2}$}
\author{Jing Qian$^{2,4}$}
\email{Corresponding author: jqian1982@gmail.com}
\author{Weiping Zhang$^{1,3,4,5}$}

\affiliation{$^{1}$School of Physics and Astronomy, and Tsung-Dao Lee Institute, Shanghai Jiao Tong University, Shanghai, 200240, China}
\affiliation{$^{2}$State Key Laboratory of Precision Spectroscopy, Department of Physics, School of Physics and Electronic Science, East China Normal University, Shanghai, 200062, China
}
\affiliation{$^{3}$Shanghai Research Center for Quantum Science, Shanghai, 201315, China}
\affiliation{$^{4}$Shanghai Branch, Hefei National Laboratory, Shanghai 201315, China}
\affiliation{$^{5}$Collaborative Innovation Center of Extreme Optics, Shanxi University, Taiyuan, Shanxi 030006, China}
%\author{Jing Qian}%
%\affiliation{%


\begin{abstract}
Rydberg quantum gate serving as an indispensable computing unit for neutral-atom quantum computation, has attracted intense research efforts for the last decade. However the state-of-the-art experiments have not reached the high gate fidelity as predicted by most theories due to the unexpected large loss remaining in Rydberg and intermediate states. In this paper 
we report our findings in constructing a native two-qubit controlled-NOT gate based on pulse optimization. We focus on the method of commonly-used two-photon Rydberg excitation with smooth Gaussian-shaped pulses which is straightforward for experimental demonstration. By utilizing optimized pulse shapes the scheme reveals a remarkable reduction in the decays from Rydberg and intermediate states, as well as a high-tolerance to the residual thermal motion of atoms. We extract a conservative lower bound for the gate fidelity \textcolor{black}{$> 0.992$} after taking into account the experimental imperfections. Our results not only reduce the gap between experimental and theoretical prediction because of the optimal control, but also facilitate the connectivity of distant atomic qubits in a larger atom array by reducing the requirement of strong blockade, which is promising for developing multiqubit quantum computation in large-scale atomic arrays.
\end{abstract}

\maketitle

%\tableofcontents

\section{\label{sec:level1}Introduction}

Trapped neutral atoms with highly-excited Rydberg states are a promising candidate for universal quantum computation because of long-time coherence and strong long-range interaction \textcolor{black}{\cite{Adams_2020,Henriet2020quantumcomputing,doi:10.1116/5.0036562,PhysRevX.12.021049,Shi_2022}}. Major advantages over other technologies are the construction of sizable neutral atom arrays in one \textcolor{black}{\cite{doi:10.1126/science.aah3752}}, two \textcolor{black}{\cite{doi:10.1126/science.aah3778,PhysRevLett.122.203601,PhysRevLett.128.083202}} and three dimensions \textcolor{black}{\cite{Barredo2018,Kumar2018}}, with which the platform is made essentially scalable. A computationally universal quantum computing set can be built on neutral-atom qubits via single- and two-qubit gate operations. Single-qubit gates with exquisite fidelity exceeding 0.99 have been demonstrated by microwave or optical transitions \textcolor{black}{\cite{Olmschenk_2010,PhysRevLett.114.100503,doi:10.1126/science.aaf2581,PhysRevLett.121.240501}}. Protocols for entangling two atoms using Rydberg interactions have also been explored \textcolor{black}{\cite{PhysRevLett.104.010502,PhysRevA.82.030306,Picken_2019,Madjarov2020,PhysRevLett.124.033603}}, however despite these significant advances, the fidelities for atomic two-qubit gates either $C_Z$ gate(0.89 \textcolor{black}{\cite{PhysRevLett.123.230501}}, 0.97 \textcolor{black}{\cite{PhysRevLett.123.170503}}) or controlled-NOT(CNOT) gate(0.72 \textcolor{black}{\cite{PhysRevLett.104.010503}}, 0.73 \textcolor{black}{\cite{PhysRevLett.119.160502}}, 0.82 \textcolor{black}{\cite{PhysRevLett.129.200501}}), remain limited to much low values as compared with other competitive platforms such as trapped ions(0.999 \textcolor{black}{\cite{PhysRevLett.117.060504}}, 0.999 \textcolor{black}{\cite{PhysRevLett.117.060505}}) or superconducting circuits(0.994 \textcolor{black}{\cite{Barends2014}}, 0.995 \textcolor{black}{\cite{PhysRevLett.125.240503}}, 0.997 \textcolor{black}{\cite{PhysRevLett.126.220502}}). The main reason for this infidelity is the unexpected loss from Rydberg and intermediate states that causes severe dephasing process affecting the Ramsey oscillation between the ground and Rydberg states \textcolor{black}{\cite{PhysRevApplied.13.024008}}. To overcome it one may adopt low-noise laser sources \cite{PhysRevLett.121.123603} or a dynamical-decoupling approach without encoding \cite{PhysRevApplied.19.034069}.

Although the theoretical limitation for an atomic two-qubit gate has extended to $>0.9999$ \textcolor{black}{\cite{PhysRevA.96.042306}} yet the experimental results reported are still far away from this value. \textcolor{black}{Until recently, a native CNOT gate proposal originally proposed by M\"{u}ller and coworkers \cite{PhysRevLett.102.170502}, is experimentally demonstrated} in which the loss corrected fidelity number is only 0.82 \textcolor{black}{\cite{PhysRevLett.129.200501}}. One of the reasons that leads to a huge gap between experimental demonstration and theoretical prediction is the unexpected atom loss remaining in the Rydberg and intermediate-excited states which is caused by the use of $\pi-\text{gap}-\pi$ pulse sequence required by a typical Rydberg blockade gate \textcolor{black}{\cite{PhysRevA.88.062337,PhysRevA.94.032306,Isenhower:11,farouk2022parallel}}. 
By comparing with this pulse method, utilizing a time-optimal continuous pulse on the ground-Rydberg transition is more straightforward 
which can avoid the need of single-site atom addressability allowing for executing more qubits natively
\textcolor{black}{\cite{Jandura2022timeoptimaltwothree,PhysRevA.89.032334,Sun:23}}.
Experimental demonstration with a globally-modulated time-optimal pulse has achieved a competing fidelity of 0.98 for a two-qubit $C_Z$ gate \textcolor{black}{\cite{PhysRevA.105.042430}}.
Nevertheless, although a large number of research efforts have been devoted, the atomic two-qubit gate fidelity is still far below 0.99 in most experiments, which poses an urgent task towards a high-fidelity gate scheme that can be readily for practical implementation.

In this work, we report a practical two-qubit CNOT gate via smoothly-modulated Gaussian pulses driving a native two-photon transition which is commonly used by experiments \textcolor{black}{\cite{PhysRevLett.100.253001,PhysRevLett.110.123001,PhysRevLett.114.113002}}. The two-qubit CNOT gates are useful because they can achieve arbitrary multiqubit gate operations while combining with some single-qubit gates \textcolor{black}{\cite{PhysRevA.52.3457}}, which are key for realizing a universal gate set. In our approach the atoms are driven by 
time-optimal laser pulses and two-photon detuning that reduce the requirement for strong Rydberg blockade. To investigate the gate performance under experimental conditions we present a detailed error model indicating that the laser amplitude and phase noise are the dominant error sources due to the use of nontrivial waveform modulation \cite{PhysRevApplied.18.044042}. We benchmark this gate by a raw fidelity to be $\mathcal{F}=0.9987$ with a 1.0-$\mu$s duration. After accounting for tremendous technical disturbances we stress the realistic gate fidelity is still higher than \textcolor{black}{0.9921} which provides a more reasonable lower bound for experimental exploration. In addition we show the prospection for achieving even faster($\sim0.1$ $\mu$s) and higher-fidelity(\textcolor{black}{$\geq0.9929$}) two-qubit CNOT gates, highlighting the potential use of quantum optimal control methods for ultrafast quantum computation \textcolor{black}{\cite{Goerz_2011,Omran2019,PhysRevLett.103.240501}}.

\section{\label{sec:level2}Theoretical Strategy}

\begin{widetext}
    
\begin{figure}%FIGURE 
\includegraphics[width=0.65\textwidth]{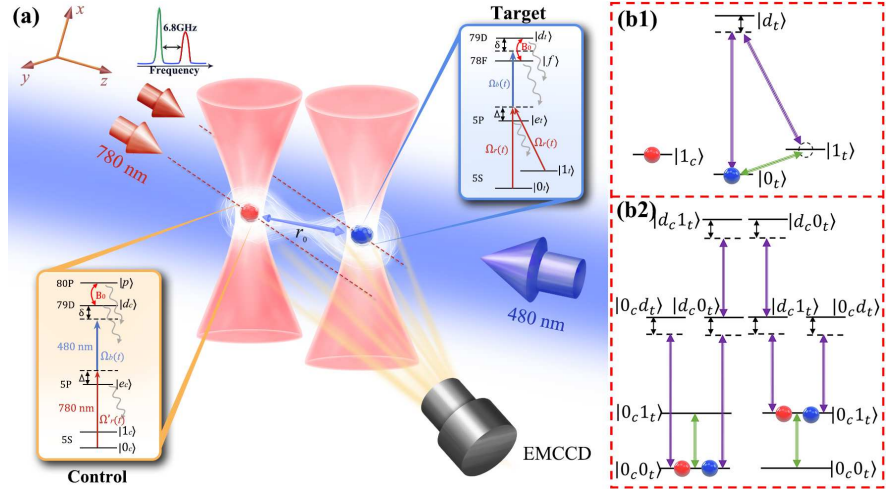} 
\caption{Experimental sketch for a two-qubit gate scheme using a native two-photon Rydberg excitation with 780 nm and 480 nm lasers in counterpropagating geometry. Control(red) and target(blue) atoms are trapped in two optical tweezers, whose energy levels are illustrated in insets. Two atoms prepared in the Rydberg pair state will experience a natural dipole-dipole exchange interaction with strength $B_0$, {\it i.e.} the two-body F\"{o}rster resonance $|d_cd_t\rangle\rightleftarrows |pf\rangle$ \textcolor{black}{\cite{PhysRevA.92.042710}}. Energy levels of single $^{87}$Rb atom are:  $|0_{c(t)}\rangle=|5S_{1/2},F=1,m_F=0\rangle$, $|1_{c(t)}\rangle=|5S_{1/2},F=2,m_F=0\rangle$, $|e_{c(t)}\rangle=|5P_{3/2}\rangle$ and $|d_{c(t)}\rangle=|79D\rangle$ \cite{PhysRevApplied.15.054020}. The dominant pair state nearby coupling to $|d_cd_t\rangle=|79D;79D\rangle$ is $|pf\rangle=|80P;78F\rangle$.
Effective models for inputs (b1) $|1_c0_t\rangle$(equivalent to $|1_c1_t\rangle$) and (b2) $|0_c0_t\rangle$(left) or $|0_c1_t\rangle$(right) under the condition of a large intermediate detuning $\Delta\gg\Omega_b,\Omega_r$.
$\Omega_{m/n}$(green/purple arrow) and $\delta_{e}$(black arrow) respectively stand for effective Rabi frequencies and effective two-photon detuning. }
\label{Fig1.model}
\end{figure}

\end{widetext}

{\it Scheme Setup.} To implement a two-qubit CNOT gate, we consider two atoms individually prepared in two optical tweezers separated by a distance $r_0$ \textcolor{black}{\cite{Urban:09}}. As illustrated in Fig.\ref{Fig1.model}(a), the qubits are encoded into hyperfine ground states $|0_{c,t}\rangle$ and $|1_{c,t}\rangle$ of $^{87}$Rb atoms which produces four computational basis states $\{|0_c0_t\rangle,|0_c1_t\rangle,|1_c0_t\rangle,|1_c1_t\rangle\}$.
 The excitation of control atom(in red) from $|0_c\rangle\to|d_c\rangle$ is driven by a two-photon transition with intermediate state $|e_c\rangle$ and coupling strengths 
$\Omega_r^\prime(t)$ and $\Omega_b(t)$. $|1_c\rangle$ is uncoupled and $\Delta$ represents the center-of-mass detuning from $|e_c\rangle$. For the target atom(in blue), the ground states $|0_t\rangle$ and $|1_t\rangle$ are off-resonantly coupled to the intermediate state $|e_t\rangle$ via an equal detuning $\Delta$ which is ensured by two phase-locked Raman lasers $\Omega_r(t)$ \textcolor{black}{\cite{PhysRevLett.129.200501}}.
A second laser couples
$|e_t\rangle\to|d_t\rangle$ with Rabi frequency $\Omega_b(t)$. 
$\delta$ treats as a (small) detuning from the two-photon resonance, which can also be desirably tuned via external fields \textcolor{black}{\cite{Ravets2014}}.
All possible spontaneous decays from lossy states including $|e_{c,t}\rangle$, $|d_{c,t}\rangle$, $|p\rangle$, $|f\rangle$ are considered, denoted by the rates $\Gamma_e$ and $\Gamma_{d,p,f}$. \textcolor{black}{The description above presents a simplified model. A more realistic case involving multiple hyperfine intermediate states $|5P_{3/2},f_e,m_{f_e}\rangle$ is discussed in Appendix \ref{appendix}.}

We propose a two-photon adiabatic method to facilitate the gate in which \textcolor{black}{the 480 nm coupling laser globally covers two optical traps as equally as possible. While the 780 nm excitation lasers $\Omega_r^\prime(t)$, $\Omega_r(t)$ independently illuminate the control and target atomic qubits. Especially the Raman laser $\Omega_r(t)$ driving $|0_t\rangle\to|1_t\rangle$
experiences electro-optic modulation to generate a sideband of $6.8$ GHz({\it i.e.} the hyperfine splitting of ground states) that can separately couple $|0_t\rangle\to|e_t\rangle$ and $|1_t\rangle\to|e_t\rangle$ with an equal amplitude. To accelerate pulse optimization we use same Rabi frequency $\Omega_r^\prime(t)=\Omega_r(t)$.}
 If an effective scheme without the intermediate states $|e_{c,t}\rangle$ is presumed the gate operation can even be performed via one excitation laser \textcolor{black}{\cite{PhysRevApplied.17.024014}}. Here we modulate all pulse waveforms to be smooth Gaussian shaping, as it is more suitable to suppress non-adiabatic leakage during the gate execution, promising for an improvement of the gate fidelity \textcolor{black}{\cite{PhysRevA.94.032306}}. Except for experimental imperfections(see detailed discussions in Sec.{V}) the population leakage intrinsically comes from two aspects: one is due to the spontaneous emission from Rydberg and intermediate states, the other is due to imperfect optimization which makes the excitation of double Rydberg states $|d_cd_t\rangle$, $|pf\rangle$ enhanced. A proper time-dependent modulation for the two-photon detuning $\delta(t)$ can strongly improve the optimization procedure as we observe in later discussions.

{\it Gate procedure.} For the case of the control atom initially in the uncoupled state $|1_c\rangle$ as shown in Fig.\ref{Fig1.model}(b1), the Hamiltonian for the target atom is ($\hbar=1$)
\begin{eqnarray}
    \mathcal{H}_{t} &=& -\Delta\left\vert e_t\right\rangle\left\langle e_t\right\vert-\delta\left\vert d_t\right\rangle\left\langle d_t\right\vert
    +\frac{\Omega_{r}}{2}(\left\vert 0_t\right\rangle \left\langle e_t\right\vert +\left\vert 1_t\right\rangle \left\langle e_t\right\vert) \nonumber\\
    &+&\frac{\Omega_{b}}{2}\left\vert e_t\right\rangle\left\langle d_t\right\vert +h.c. 
\end{eqnarray}
which, if $|\Delta|\gg \Omega_r,\Omega_b$, allows adiabatic elimination of the intermediate state $|e_t\rangle$. That leads to a reduced closed-loop model for the target atom with effective couplings and detuning described by \textcolor{black}{\cite{Brion_2007}}
\begin{equation}
 \Omega_n = \frac{\Omega_b \Omega_r}{2\Delta},\Omega_m=\frac{\Omega_r^2}{2\Delta},\delta_{e}=\frac{\Omega_r^2-\Omega_b^2}{4\Delta}+\delta,
 \label{dd}
\end{equation}
As seen in Fig.\ref{Fig1.model}(b1), 
the combination of $\Omega_n$ and $\Omega_m$ can swap the population between $|0_t\rangle$ and $|1_t\rangle$, arising
\begin{equation}
|1_c0_t\rangle\rightleftarrows|1_c1_t\rangle .
 \label{tr1}
\end{equation}

If instead, the control atom is in state $|0_c\rangle$, the
Hamiltonian for it reads
\begin{eqnarray}
   {\mathcal{H}_{c}}&=& -\Delta\left\vert e_c\right\rangle\left\langle e_c\right\vert-\delta\left\vert d_c\right\rangle\left\langle d_c\right\vert+\frac{\Omega_{r}}{2}\left\vert 0_c\right\rangle \left\langle e_c\right\vert \nonumber \\
    &+&\frac{\Omega_b}{2}\left\vert e_c\right\rangle \left\langle d_c\right\vert 
    +h.c. 
\end{eqnarray}
and a coherent coupling between $|0_c\rangle$ and $|d_c\rangle$ with effective Rabi frequency $\Omega_n$ and detuning $\delta_{e}$, can be established. Fig.\ref{Fig1.model}(b2) represents the whole diagram of the two-atom picture including relevant energy levels and effective couplings. Note that state $|d_c d_t\rangle$ which
induces a resonant dipole-dipole shift $B_0$ described by 
\begin{equation}
  \mathcal{H}_{dd} = B_0(|d_cd_t\rangle\langle pf|+|pf\rangle\langle d_cd_t|)
\end{equation}
has been ignored in the figure. Because when $B_0\gg\Omega_m,\Omega_n$ it is sufficient to block the double Rydberg excitation making $|d_cd_t\rangle$ and $|pf\rangle$ rarely unoccupied. The exchange interaction energy is appropriately $B_0=C_3/r_0^3$ where $C_3/2\pi = 23.276$ GHz$\cdot\mu$m$^3$ given by the ARC library \textcolor{black}{\cite{SIBALIC2017319}}. This interaction energy shift results in a desired state transformation that obeys ($\beta\in \{|0_t\rangle,|1_t\rangle\}$)
\begin{equation}
    |0_c\beta\rangle\to|0_c\beta\rangle.
\end{equation}

To this end, our protocol can principally provide a direct realization of a natural two-qubit CNOT gate without additional single-qubit rotations. However, 
 the main difficulty for practical implementation lies in finding accurate Rabi frequencies which lead to ideal state transformation. 
Since time-optimal laser pulses are prospective for a perfect adiabatic evolution, we first perform careful optimization to the amplitude of
two native laser pulses \textcolor{black}{\cite{Pelegri_2022}}, which can be readily accessible for experimental demonstration.
This two-photon Rydberg excitation configuration offers extra modulations for the coupling laser and the two-photon detuning which 
significantly differs from the single-photon transition with one optimal laser \textcolor{black}{\cite{Han_2016,PhysRevApplied.13.024059}}.

\vspace{20pt}
\section{Pulse Design and Gate performance}

\begin{widetext}

\begin{figure}%FIGURE 
\includegraphics[width=0.7\textwidth]{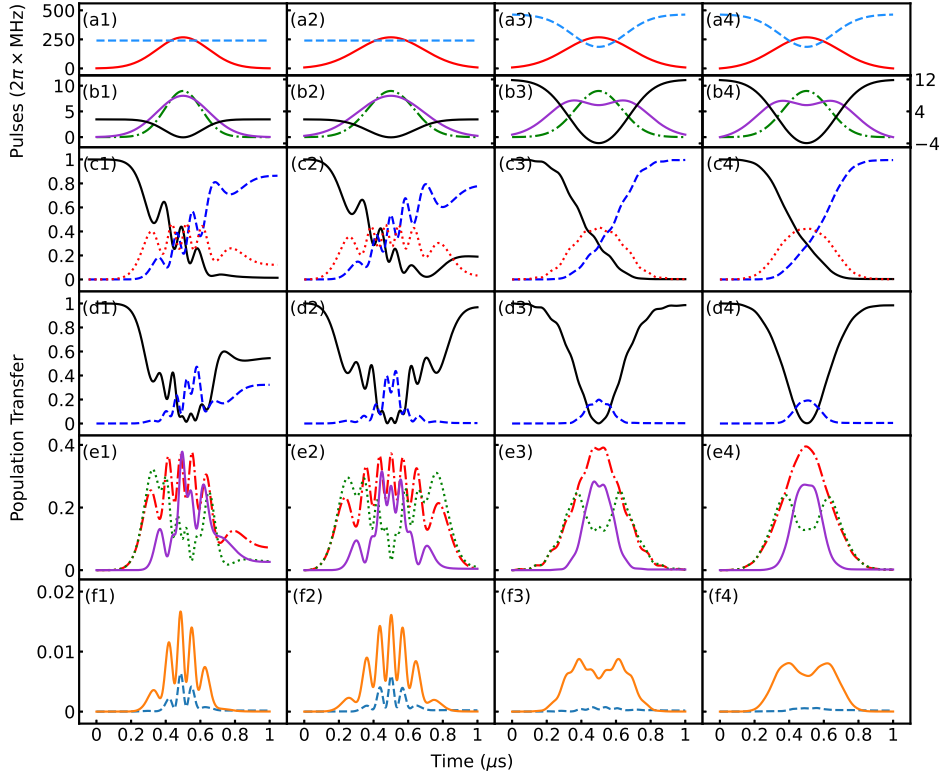} 
\caption{Performance of the two-photon CNOT gates with optimal Gaussian pulses.
Time dependence of (a1) the laser Rabi frequencies $\Omega_r(t)$(red-solid), $\Omega_b(t)$(blue-dashed), of (b1) the effective Rabi frequencies $\Omega_m$(green-dash-dotted), $\Omega_n$(purple-solid) and the effective detuning $\delta_e(t)$(black-solid). (c1-f1) The time evolution of state population.
In (c1) states $|1_c0_t\rangle$(or $|1_c1_t\rangle$), $|1_cd_t\rangle$, $|1_c1_t\rangle$(or $|1_c0_t\rangle$) are represented by the black-solid, red-dotted, blue-dashed lines, corresponding to the initial states of $|1_c0_t\rangle$(or $|1_c1_t\rangle$). In (d1-f1) states $|0_c0_t\rangle$(or $|0_c1_t\rangle$), $|0_c1_t\rangle$(or $|0_c0_t\rangle$), $|0_cd_t\rangle$, $|d_c0_t\rangle$(or $|d_c1_t\rangle$), $|d_c1_t\rangle$(or $|d_c0_t\rangle$), $|d_cd_t\rangle$, $|pf\rangle$ are represented by black-solid, blue-dashed, red-dash-dotted, green-dotted, purple-solid, orange-solid, blue-dashed lines, corresponding to the initial states of $|0_c0_t\rangle$(or $|0_c1_t\rangle$).
Population of state $|pf\rangle$ is magnified by a factor of 10 in (f1).
Similarly, different Gaussian pulses are applied in (a2-f2), (a3-f3), (a4-f4).
Pulse parameters are described in the main text and other common parameters are $B_0/2\pi=32$ MHz for a $r_0= 9$ $\mu$m separation, $\Delta/2\pi = 4.0$ GHz, $
\delta = 10$ MHz, $\Gamma_e/2\pi = 3$ MHz, $\Gamma_{d,p,f}/2\pi = 3$ kHz.} %Subscripts {\it c,t} are omitted for the sake of convenience.}
\label{Fig2.compare}
\end{figure}

\end{widetext}

To simulate the gate, we choose smoothly-modulated Gaussian pulses and then perform careful optimization of their shapes. The advantage of Gaussian modulation lies in its better feasibility of the shape to be produced \textcolor{black}{\cite{PhysRevResearch.4.033019}}
as compared with other complex pulse waveforms \textcolor{black}{\cite{PhysRevA.101.062309,PhysRevA.90.032329,PhysRevA.84.042315}}.

In a first approach, \textcolor{black}{as done by \cite{PhysRevA.105.042430}} we assume a Gaussian waveform individually for the 780 nm laser
\begin{equation}
\Omega_r(t) = \Omega_r^{\max} e^{-\frac{(t-T_g/2)^2}{2w_r^2}}
\end{equation}
and keeping
\begin{equation}
    \Omega_b=\Omega_b^{\max}(\text{constant})
\end{equation}
for the 480 nm laser, in which $\Omega_{r,b}^{\max}$, $T_g$, $w_r$ are the peak amplitudes, the gate time, the pulse width, respectively. 
We first calculate the dynamics of system driven by a single Gaussian pulse $\Omega_r(t)$ where $\Omega_r^{\max}/2\pi=268$ MHz, $\Omega_b/2\pi=240$ MHz and $\omega_r=0.15$ $\mu$s are arbitrarily chosen. 
 By evolving the master equation (\ref{raa}) [see numerical method in Appendix \ref{appendixa}],
 in Fig. \ref{Fig2.compare}(a1-b1) we present the time-dependence of laser pulses $\Omega_r(t)$ and $\Omega_b$, as well as the effective Rabi frequencies $\Omega_{m}(t)$, $\Omega_{n}(t)$ and detuning $\delta_e(t)$. The corresponding state evolution is shown in (c1-f1). Based on such pair of non-optimal laser pulses we observe the presence of strong dramatic oscillations in the population transfer due to the off-resonant transitions among undesired intermediate states. The resulting gate fidelity, accounting for the effect of four computational basis states,
 is only $\mathcal{F}=0.8338$. 
 So we proceed to perform optimization to the pulse width $\omega_r$ via genetic algorithm towards the target of maximizing $\mathcal{F}$. The mechanism of 
algorithm can be found in our prior work \textcolor{black}{\cite{PhysRevApplied.17.024014}}.
Results in Fig. \ref{Fig2.compare}(a2-f2) show that, with an optimal $\omega_r=0.1899$ $\mu$s, the effective Rabi frequencies $\Omega_{n}$ and $\Omega_{m}$ are broadened in the time domain which apparently makes the transfer efficiency enhanced. The calculated gate fidelity reaches $\mathcal{F}=0.9329$. 

However, we realize the way of single-Gaussian-pulse optimization can not solve the problem of strong oscillations in the population transfer. Because $\Omega_b(t)$ is kept constant during the total gate execution that breaks the adiabaticity of system \textcolor{black}{\cite{PhysRevA.89.030301,PhysRevA.96.022321,Yu:19,PhysRevX.10.021054}}. Therefore, in a second approach, we demonstrate the gate incorporating with another Gaussian modulation for $\Omega_b(t)$ (using $\Omega_r(t)$ as in case (a2)), 
\begin{equation}
    \Omega_b(t)=\Omega_{bm}e^{-\frac{(t-T_g/2)^2}{2w_b^2}}+K
\end{equation}
where $\Omega_{bm}$, $\omega_b$, $K$ denote the amplitude, the pulse width and the light shift. After optimization we obtain a set of satisfactory parameters $(\Omega_{bm}/2\pi,\omega_b,K/2\pi)=(-279.2$ MHz, 0.1546 $\mu s$, 464.9 MHz$)$, with which
$\Omega_n(t)$ and $\delta_e(t)$ are modulated more sufficiently, increasing the adiabaticity of protocol.
Fig.\ref{Fig2.compare}(c3-f3) depict the dynamics of each input state in detail which confirms that the population transfer becomes smoother
and higher-efficiency accompanied by a suppression of the intermediate-state population.
After averaging over four computational states, we find the gate fidelity can reach $0.9955$. 
Since the Gaussian function may suffer from extra disadvantage due to the nonvanishing tail at the start and the end we also make a correction on the standard pulse by employing
\begin{equation}
\Omega_r^{corr}(t) = \Omega_r^{\max}[ \frac{e^{-{(t-T_g/2)^2}/{2w_r^2}}-e^{-{(T_g/2)^2}/{2w_r^2}}}{1-e^{-{(T_g/2)^2}/{2w_r^2}}}]
\end{equation}
instead of $\Omega_r(t)$ and keep $\Omega_b(t)$ unchanged. The corresponding gate fidelity becomes 0.9948 which means the error caused by a nonvanishing tail of Gaussian pulses is not important. Nevertheless the smoothness of population transfer is dramatically improved in the fourth case, indicating perfect adiabaticity preserved during the gate operation.

In the second approach with a dual-Gaussian-pulse optimization we also observe that the intermediate singly-excited Rydberg states are heavily populated with a peak amplitude about $\sim 0.4$(see {\it e.g.} Fig.\ref{Fig2.compare}(c4),(e4)) in the adiabatic evolution. As a consequence, 
the intrinsic spontaneous-decay error due to the finite lifetime of energy levels plays a major role. We numerically find a decay error of Rydberg(or intermediate) levels is $3.1\times10^{-3}$(or $1.1\times10^{-3}$). In principle this decay error sets a theoretical limitation for the ultimate gate fidelity apart from experimental imperfections \textcolor{black}{\cite{PhysRevA.97.053803}}. 
One way to overcome it resorts to reduce the near-resonance excitation onto intermediate singly-excited Rydberg states by tuning the effective detuning $\delta_e$ off-resonance \textcolor{black}{\cite{PhysRevA.88.010303}}. With a constant detuning $\delta=10$ MHz as used in calculating Fig.\ref{Fig2.compare}, the effective $\delta_e(t)$ value(black-solid lines in (b1-b4)) will experience two-photon resonance when $\delta_e(t)=0$, leading to an enhanced Rydberg excitation accompanied by a relatively large spontaneous decay. A time-dependent optimization to the bare $\delta$ value may reduce the influence of spontaneous emission through suppressing the population onto these undesired excited states.

\vspace{10pt}
\section{Detuning optimization}

\begin{table*}
\caption{\label{tab:table3}{(i-iii) Coefficients of the optimal laser pulses $\Omega_r(t),\Omega_b(t)$ and the detuning $\delta(t)$ under different gate durations $T_g=(1.0,0.25,0.1)$ $\mu$s. The raw gate fidelity $\mathcal{F}$ is given accordingly. No external fluctuation is considered.}}
\begin{ruledtabular}
\setlength{\tabcolsep}{0mm}{
\begin{tabular}{ccccccccccc}
$\quad$&\multicolumn{1}{c}{$T_g$($\mu$s)}&\multicolumn{2}{c}{$\Omega_r(t)/2\pi$}&\multicolumn{3}{c}{$\Omega_b(t)/2\pi$} &\multicolumn{3}{c}{$\delta(t)/2\pi$}&$\mathcal{F}$\\
\hline
$\quad$&$\quad$&$\Omega^{max}_r$(MHz)&$\omega_r$($\mu$s)&$\Omega_{bm}$(MHz)&$\omega_b$($\mu$s)&$K$(MHz)&$\delta_0$(MHz)&$\delta_1$(MHz)&$\delta_2$(MHz)&$\quad$\\
%\cmidrule(r){2-3}\cmidrule(r){4-6}\cmidrule(r){7-9}
\hline
(i)&1.0&268&0.0921&538.01&0.9885&136.47&-11.33&-42.38&-18.71&0.9987\\
\hline
(ii)&0.25&268&0.2828&426.36&0.2796&296.82&-54.79&15.56&88.22&0.9986\\
\hline
(iii)&0.1&306&0.8045&419.98&0.7336&555.05&-58.49&18.64&119.08&0.9984\\
\end{tabular}
}
\end{ruledtabular}
\end{table*}

\begin{figure}
    \includegraphics[width=0.47\textwidth]{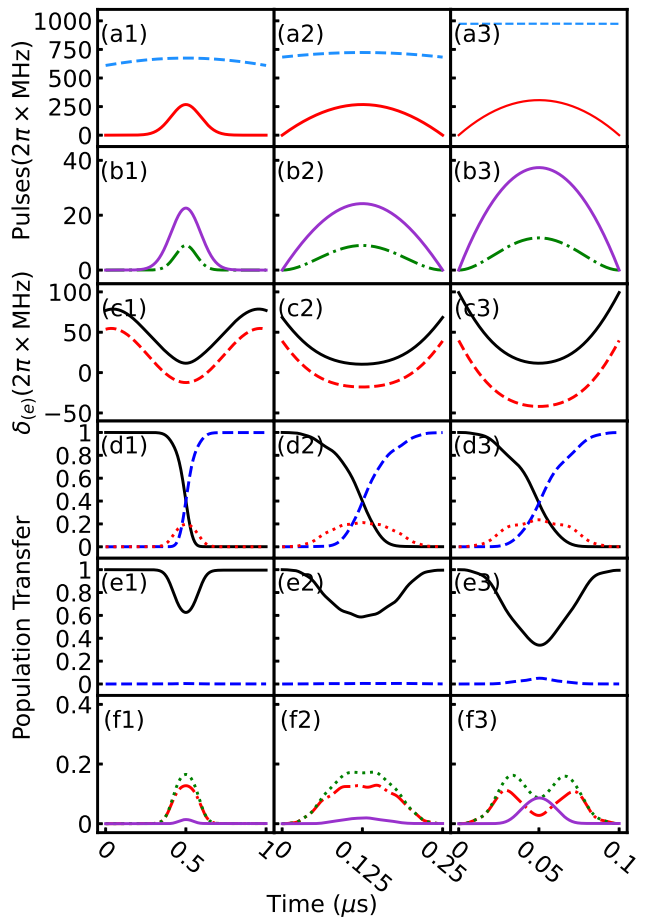} 
    \caption{Improved two-photon CNOT gates enabled by a simultaneous modulation of the two-photon detuning. From left to right the gate time is set to be $T_g=(1, 0.25, 0.1)$ $\mu$s. Time dependence of (a1-a3) laser Rabi frequencies $\Omega_r(t)$, $\Omega_b(t)$, of (b1-b3) effective Rabi frequencies $\Omega_m(t)$, $\Omega_n(t)$, and of (c1-c3) bare detuning $\delta(t)$(red-dashed) and effective detuning $\delta_e(t)$(black-solid), is shown. (d1-f3) Time-dependent population evolution. All linetypes are same as in  Fig.\ref{Fig2.compare}.}
    \label{Fig3.improvement}
\end{figure}

Furthermore, we seek an appropriate waveform for the bare two-photon detuning $\delta(t)$ which is described by
\begin{equation}
    \delta(t)=\delta_0 + \delta_1\cos(\frac{2\pi t}{T_g})+\delta_2\sin(\frac{\pi t}{T_g})
\end{equation}
and globally re-optimize all pulse coefficients. 
After performing sufficient runs of optimization we obtain sets of satisfactory parameters $(\Omega_r(t),\Omega_b(t),\delta(t))$ for different $T_g$ values as summarized in Table.\ref{tab:table3}. Detailed population evolutions are given in Fig.\ref{Fig3.improvement}.
In general, benefiting from the use of an optimized two-photon detuning $\delta(t)$(red-dashed) in (c1-c3), 
the effective detuning $\delta_e(t)$(black-solid) is always positive and kept far away from resonance, strongly reducing  the population remain in the Rydberg state.  
The peak amplitude of singly-excited Rydberg states has been suppressed to be lower than $0.2$ during the adiabatic evolution. Therefore, the intrinsic infidelity loss which is primarily limited by the finite lifetime of intermediate and Rydberg energy levels, can be deeply suppressed.
We find a higher gate fidelity $\mathcal{F}>0.998$(see the last column in Table I) for all cases (i-iii). 
The decay error in case (i) is only $6.44\times 10^{-4}$($6.24\times 10^{-4}$) coming from the spontaneous decay of Rydberg(intermediate) states, which is smaller than that in the case of a constant detuning $\delta$ by
one order of magnitude. More strikingly, other intrinsic errors coming from imperfect optimization, finite Rydberg interaction strength(see Sec.\ref{subsectionc}) and rotation due to the dipole-dipole exchange interaction, have been suppressed to a negligible level $\sim 10^{-5}$.

Another important advantage of detuning optimization lies in the ability to achieve a faster gate operation with a same high level of fidelity. As presented in Fig.\ref{Fig3.improvement} by decreasing $T_g$ from 1 $\mu$s to 0.1 $\mu$s the gate fidelity preserves to be larger than 0.998 at the expense of a slightly stronger coupling laser which makes the effective coupling $\Omega_n$ much stronger. A larger $\Omega_n$ leads to an enhanced population transfer, which ensures the entire operation can be accomplished within a sub-microsecond time scale $T_g\sim 0.1$ $\mu$s. The bare detuning $\delta(t)$(red-dashed in (c3)) also acquires a more sufficient modulation for keeping $\delta_e(t)$(black-solid) far from resonance, in order to avoid the excitation of singly-excited Rydberg states.
Therefore we believe that, our approach which combines the optimal control of two native laser pulses as well as a reasonable modulation for the two-photon detuning, 
can avoid the requirement for a strong Rydberg blockade, so as to reduce the gate duration of $\sim 1.0$ $\mu$s as required by a typical blockade gate \textcolor{black}{\cite{PhysRevA.82.030306}}. It is promising for realizing a fast quantum gate (see Extended study Appendix \ref{appendixb}).

\section{Gate Robustness under practical conditions}

Any physical realization of quantum gates will inevitably deviate from the ideal scenario due to experimental imperfections.
The source of these imperfections can be the atomic motion under a finite temperature, the technical noise from excitation lasers and a finite interaction strength caused by external fields. This section is dedicated to analyze the effect of these three types of imperfections on the gate performance. Here we adopt case (i) $T_g=1.0$ $\mu$s.

\subsection{Influence from finite atomic temperature}

\begin{figure}%FIGURE 
\includegraphics[width=0.48\textwidth]{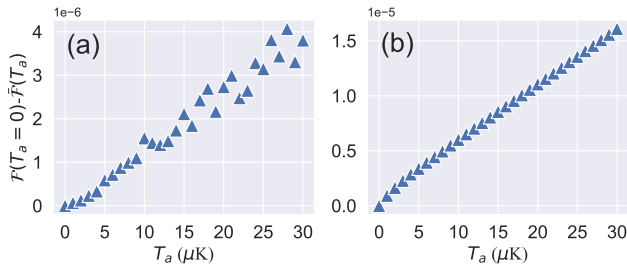} 
\caption{Infidelity caused by the residual thermal motion of cold atoms under different temperatures $T_a \in [0,30]$ $\mu$K. (a) Position error from a fluctuated interaction between two trapped atoms. (b) Doppler dephasing error due to the laser frequency shift felt by atoms. Each point denotes an average of $\mathcal{N}=500$ stochastic simulations.} 
\label{Fig4.temperature}
\end{figure}

To our knowledge, the temperature of atoms cannot reach absolute zero at the present stage. The residual thermal motion of atoms is an important error factor which affects the gate performance \textcolor{black}{\cite{PhysRevLett.123.230501,Wu:21}}.

{\it Fluctuated atomic position -} First of all, a nonzero temperature will lead to the uncertainty of atom position in the optical traps and thereby to a fluctuated interatomic interaction. A scheme usually relying on strong blockade \textcolor{black}{\cite{PhysRevLett.123.170503}} or using large distances \textcolor{black}{\cite{PhysRevA.94.062307}}, could tolerate this position error. 
Here, accounting for the thermal motion of trapped atoms we consider the atomic position distribution given by a 3D Gaussian function:
\begin{equation}
    f(\boldsymbol{r})=\frac{1}{(2\pi)^{3/2}\sigma_x\sigma_y\sigma_{z}}e^{-\frac{x^2}{2\sigma_x^2}}e^{-\frac{y^2}{2\sigma_y^2}}e^{-\frac{z^2}{2\sigma_z^2}} \label{3D}
\end{equation}
where the standard deviation $\sigma_{x,y,z}=\sqrt{k_B T_a/m\omega_{x,y,z}^2}$ depends on the atomic temperature $T_a$ and trapping frequencys $\omega_{x,y,z}$. $k_B$ is the Boltzmann constant and $m$ is the atomic mass. Since the experiments with Rydberg atoms in optical tweezers are usually performed with hot atoms we tune the temperature in the range of $T_a \in [0,30]$ $\mu$K and calculate the infidelity $\mathcal{F}(T_a=0)-\bar{\mathcal{F}}(T_a)$ with a fluctuated interaction ${B}(\boldsymbol{r}_j)$. The uncertainty in atomic position leads to a fluctuated dipole-dipole interaction, given by
\begin{equation}
    B(\boldsymbol{r}_{j})\approx B_0+\delta B(\boldsymbol{r}_{j}-\boldsymbol{r}_0)
\end{equation}
where $\boldsymbol{r}_j$ is a random relative distance for each measurement $j$ and $B_0=C_3/r_0^3$ with $r_0=|\boldsymbol{r}_0|$. The uncertainty of interaction takes form of
\begin{equation}
    \delta B(\boldsymbol{r}_{i}-\boldsymbol{r}_0)=-\frac{3C_3(\boldsymbol{r}_{i}-\boldsymbol{r}_0)}{|\boldsymbol{r}_0| ^4}.
\end{equation}

During each measurement we can obtain a fidelity value $\mathcal{F}_{j}$ based on one random $B(\boldsymbol{r}_j)$, and the ultimate estimation for a practical CNOT gate depends on an average over sufficient measurements
\begin{equation}
    \bar{\mathcal{F}}=\frac{1}{\mathcal{N}}\sum_{j=1}^{\mathcal{N}}{\mathcal{F}_{j}}
\end{equation}
where all optimized pulse parameters are adopted
as in the non-fluctuated case (i) of Table I.
The numerical results shown in Fig.\ref{Fig4.temperature}(a) reveal that the average position error ${\mathcal{F}}(T_a=0)-\bar{\mathcal{F}}(T_a)$ increases {\it vs} the atomic temperature $T_a$. Nevertheless, this value keeps at a negligible level $\leq 4\times 10^{-6}$ as compared to the intrinsic decay error($\sim 10^{-4}$). Because for hot atoms ($T_a=30$ $\mu$K) with large fluctuations the maximal uncertainty of atomic position is $\sigma_{x,y,z}=(60,70,250)$ nm calculated by using trap frequencies $\omega_{x,y,z}/2\pi=(147,117,35)$ kHz \textcolor{black}{\cite{Chew2022}}, arising a relatively small uncertainty of interaction $\delta B/B_0 \approx 8.75\%$. This result verifies the robustness of our gate scheme against position error from residual thermal motion of atoms, and a high gate fidelity can be strongly preserved with hotter atoms.

{\it Doppler effect -} Another imperfection caused by finite atomic temperature is the Doppler effect. The ground-Rydberg excitation suffers from atomic motional dephasing, which makes the real laser frequency perceived by a moving atom deviate from the original value. This can be estimated by a replacement of $\Omega_{b,r} \rightarrow \Omega_{b,r} e^{i\Delta_{b,r} t}$ where $\Delta_{b,r}$ is an extra phase factor of the Rabi frequencies and should meet a 1D Gaussian distribution:
\begin{equation}
    f(\Delta_{b,r})=\frac{1}{\sqrt{2\pi}\sigma_{\Delta_{b,r}}}e^{-\frac{(\Delta_{b,r}-\bar{\Delta}_{b,r})^2}{2\sigma_{\Delta_{b,r}}^2}}
    \label{gg}
\end{equation}
with respect to the mean $\bar{\Delta}_{b,r}=0$ and the standard deviation $\sigma_{\Delta_{b,r}}=\vec{{k}}_{b,r}\cdot\vec{{v}}_{rms}$. This implies, for each realization, the real laser detuning felt by atoms is a random variable. Here $\vec{k}_{b,r}$ is the laser wavevector and $|\vec{v}_{rms}|=\sqrt{k_BT_a/M}$ is the 1D rms speed, which lead to the standard deviations $\sigma_{\Delta_{b}}=k_b{v_{rms}}\approx 704.4$ kHz, $\sigma_{\Delta_{r}}=k_r{v_{rms}}\approx 433.4$ kHz at $T_a=30$ $\mu$K. To reduce the motional dephasing effect, we assume the excitation lasers $\Omega_{b}(t)$ and $\Omega_{r}(t)$ are counterpropagating as displayed by Fig.\ref{Fig1.model}(a), which make the extra laser detuning $\Delta_b$, $\Delta_r$ having opposite signs for each realization. Fig.\ref{Fig4.temperature}(b) shows the calculated dephasing error under the influence of Doppler effect with various temperatures.
 For a given $T_a$, we randomly adopt $\Delta_{b}$, $\Delta_{r}$ from function $f(\Delta_{b,r})$ to present the phase factor of Rabi frequencies, which can be translated into the detuning of energy levels. 
 From the numerical results, we find that the Doppler dephasing error is only $1.61\times 10^{-5}$ at $T_a=30$ $\mu$K which plays an ignorable role on the raw gate fidelity. The reason for this tiny dephasing error mainly arises from the use of counterpropagaing lasers which can minimize the effective energy shift described by $({k}_b-{k}_r)v_{rms}\approx 0.271$ MHz acting on the two-photon detuning. Only if we utilize copropagating excitation lasers, the dephasing error will increase to 2.6 $\times 10^{-4}$ as a consequence of a larger uncertainty $(k_b+k_r)v_{rms}\approx 1.1378$ MHz for the two-photon detuning.

 Therefore we believe this protocol
 has sufficient insensitivity to the variation of trapped atomic temperatures and is more appropriate for an experimental demonstration.

{\it Inhomogeneity in Rabi frequencies -} 
Due to a finite laser beam waist, a slight change of the atomic position {\it e.g.} caused by the thermal motion, will make the laser intensity perceived by the atoms deviate from the ideal value, arising a position-dependent Rabi frequency. This error is minimized if the laser intensity can be shaped into flat tophat profiles by SLMs in experiment \cite{Ebadi2021}.
According to Ref.\textcolor{black}{\cite{PhysRevA.85.042310}} we replace $\Omega_r(t)$, $\Omega_b(t)$ by position-dependent values $\Omega_r(t,\boldsymbol{r})$, $\Omega_b(t,\boldsymbol{r})$ which are
\begin{equation}
    \Omega_r(t,\boldsymbol{r})=\Omega_r(t,0)\frac{e^{-x^2/\omega^2_{x,r}(1+z^2/L^2_{x,r})-y^2/\omega^2_{y,r}(1+z^2/L^2_{y,r})}}{\left[(1+z^2/L^2_{x,r})(1+z^2/L^2_{y,r})\right]^{1/4}}
\end{equation}
\begin{equation}
    \Omega_b(t,\boldsymbol{r})=\Omega_b(t,0)\frac{e^{-x^2/\omega^2_{x,b}(1+z^2/L^2_{x,b})-y^2/\omega^2_{y,b}(1+z^2/L^2_{y,b})}}{\left[(1+z^2/L^2_{x,b})(1+z^2/L^2_{y,b})\right]^{1/4}}
\end{equation}
where $\boldsymbol{r}=(x,y,z)$ is the atomic position at any time $t$ satisfying the distribution function (\ref{3D}) and $\Omega_{r(b)}(t,0)$ is the original optimized waveform.
Because two traps are vertically placed with respect to the laser beams(see Fig.\ref{Fig1.model}a), atoms in the center can be uniformly illuminated by both excitation lasers \textcolor{black}{\cite{PhysRevLett.110.263201}}.
When 
$\omega_{x(y),r}=7.8$ $\mu$m, $\omega_{x(y),b}=8.3$ $\mu$m
arising the Rayleigh lengths $L_{x(y),r}=245.04$ $\mu$m, $L_{x(y),b}=450.88$ $\mu$m, the infidelity caused by inhomogeneity in Rabi frequencies is only $2 \times 10^{-7}$ for $T_a=30$ $\mu$K. And this value remains as small as $2.8 \times 10^{-6}$ when reducing the beam waists to $\omega_{x(y),r}=\omega_{x(y),b}=3.0$ $\mu$m. This implies, although our protocol is relatively sensitive to the intensity and phase of excitation lasers(see Sec.\ref{subsectionb}) the influence of position-dependent Rabi frequencies can be completely suppressed.

\subsection{Influence from excitation pulses}
\label{subsectionb}

\begin{figure}%FIGURE 
\includegraphics[width=0.49\textwidth]{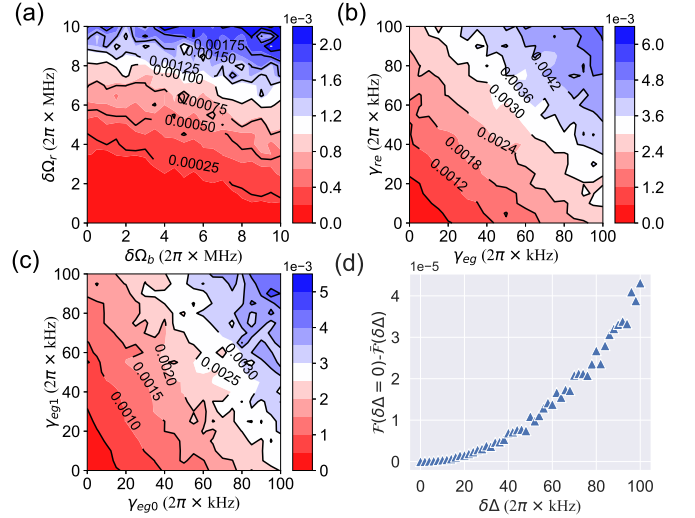} 
\caption{Infidelity caused by the fluctuations of (a) laser intensities, finite laser phase noises (b) between probe and coupling lasers and \textcolor{black}{(c) between two Raman lasers},
(d) two-photon detuning. 
Here $\delta\Omega_{r}$, $\delta\Omega_{b}$, $\delta\Delta$ in (a,d) stand for the fluctuation amplitudes. $\gamma_{re}$, $\gamma_{eg}$ in (b) mean the dephasing rates from Rydberg and intermediate states. \textcolor{black}{$\gamma_{eg0}$ and $\gamma_{eg1}$ in (c) present the individual dephasing rates in Raman transition for the target qubit.}
Each point is obtained by an average of 500 random samplings.}
\label{Fig6.flurabi}
\end{figure}

Another technical error probably leading to a poor gate fidelity, is the fact that the excitation laser pulses $\Omega_{r}(t),\Omega_{b}(t)$ and the two-photon detuning pulse $\delta(t)$ have inevitable noises during a practical demonstration. We will estimate the influence of these fluctuations in this subsection.

{\it Intensity fluctuation of excitation lasers -} In general, a gate scheme that depends on the numerical optimization of laser waveforms, greatly requires the accuracy of the laser intensity. A precise modulation is crucial to the success of scheme. However during the practical demonstration the intensity fluctuation of excitation lasers will introduce a deviation on the original optimized waveforms, arising
\begin{equation}
    \Omega_{r(b)}(t)\to \Omega_{r(b)}(t)+ \delta{\Omega_{r(b)}}
\end{equation}
where we treat $\delta{\Omega_{r(b)}}$ as a small fluctuation amplitude that varies in the range of $\delta{\Omega_{r(b)}}/2\pi\in[0,10]$ MHz. For each realization, the laser intensity is fluctuated by a random value obtained from $[-\delta{\Omega_{r(b)}},+\delta{\Omega_{r(b)}}]$
while the pulse shape keeps unchanged. 
Fig.\ref{Fig6.flurabi}(a) shows the dependence of the gate infidelity on the laser intensity fluctuation by averaging over sufficient measurements. As compared with $\delta\Omega_b$, it is evident that increasing $\delta\Omega_r$ reveals a stronger intensity error. When $\delta\Omega_r/2\pi=10$ MHz and $\delta\Omega_b$ is arbitrary, the laser intensity error reaches $>2.0\times 10^{-3}$. Because the gate is more sensitive to the weak probe laser that directly connects the ground and intermediate excited states. While the fluctuation of the coupling laser is relatively insensitive due to its strong power.

{\it Laser phase noise -} Another important effect which leads to dephasing of laser Rabi oscillations, arises from the laser phase noise due to different frequencies involved in the excitation lasers \textcolor{black}{\cite{PhysRevA.101.043421}}. An exact treatment of this effect should consider a phase factor which is directly related to the laser Rabi freuquency $\Omega_{r(b)}(t)$
\begin{equation}
\Omega_{r(b)}(t)\to\Omega_{r(b)}(t)\exp\left[i\phi_{r(b)}(t)\right]  
\end{equation}
in which their phases $\phi_{r(b)}(t)$ are random variables which can be \textcolor{black}{characterized by a phase noise spectral density $S_{\phi_{r(b)}}(f)$ as a function of the Fourier frequency \textcolor{black}{\cite{riehle2006frequency,PhysRevA.107.042611}}. Because $S_{\phi_{r(b)}}(f)$ essentially comes from the frequency noise distribution that is the Fourier transform of the laser line shape, a broader bandwidth of the lasers will lead to a larger phase fluctuation \cite{DiDomenico:10}. The resulting average gate fidelity will drop considerably as the bandwidth increases [see {\it e.g.} Ref.\cite{Pelegri_2022}, Fig.4c].}
Since the function $S_{\phi_{r(b)}}(f)$ is typically difficult to measure experimentally here we utilize a global dephasing model to quantify the average effect of laser phase noise which is described by an extra Lindblad superoperator $\mathcal{L}_{d}[{\rho}]$ in the master equation (\ref{raa}) \textcolor{black}{\cite{PhysRevA.99.043404}}
\begin{equation}
    \mathcal{L}_{d}[{\rho}] =  \sum_{j=1}^{2}[L_{dj}{\rho}L_{dj}^{\dagger}-\frac{1}{2}(L_{dj}^{{\dagger}}L_{dj}{\rho}+{\rho} L_{dj}^{{\dagger}}L_{dj})]
    \label{Ld}
\end{equation}
where
\begin{eqnarray}
    L_{d1}&=&\sqrt{\frac{\gamma_{re}}{2}}(\left\vert d_c\right\rangle\left\langle d_c\right\vert-\left\vert e_c\right\rangle \left\langle e_c\right\vert) + \sqrt{\frac{\gamma_{eg0}}{2}}(\left\vert e_c\right\rangle \left\langle e_c\right\vert \nonumber \\
    &-&\left\vert 0_c\right\rangle \left\langle 0_c\right\vert) \nonumber \\
    L_{d2}&=&\sqrt{\frac{\gamma_{re}}{2}}(\left\vert d_t\right\rangle \left\langle d_t\right\vert - \left\vert e_t\right\rangle \left\langle e_t\right\vert) + \sqrt{\frac{\gamma_{eg0}}{2}}(\left\vert e_t\right\rangle \left\langle e_t\right\vert \nonumber \\
    &-&\left\vert 0_t\right\rangle \left\langle 0_t\right\vert)+\sqrt{\frac{\gamma_{eg1}}{2}}(\left\vert e_t\right\rangle \left\langle e_t\right\vert-\left\vert 1_t\right\rangle \left\langle 1_t\right\vert)
\label{dephasing}
\end{eqnarray}
% \begin{eqnarray}
%     L_{d1}&=&\sqrt{\frac{\gamma_{re}}{2}}(\left\vert d_c\right\rangle\left\langle d_c\right\vert-\left\vert e_c\right\rangle \left\langle e_c\right\vert) + \sqrt{\frac{\gamma_{eg}}{2}}(\left\vert e_c\right\rangle \left\langle e_c\right\vert \nonumber \\
%     &-&\left\vert 0_c\right\rangle \left\langle 0_c\right\vert) \nonumber \\
%     L_{d2}&=&\sqrt{\frac{\gamma_{re}}{2}}(\left\vert d_t\right\rangle \left\langle d_t\right\vert - \left\vert e_t\right\rangle \left\langle e_t\right\vert) + \sqrt{\frac{\gamma_{eg}}{2}}(\left\vert e_t\right\rangle \left\langle e_t\right\vert \nonumber \\
%     &-&\left\vert 0_t\right\rangle \left\langle 0_t\right\vert-\left\vert 1_t\right\rangle \left\langle 1_t\right\vert)
% \label{dephasing}
% \end{eqnarray}
represent dephasings of the control and target atoms. $\gamma_{re}$
mean the dephasing rate caused by the phase noise of the common laser $\Omega_b(t)$. \textcolor{black}{$\gamma_{eg0}$ and $\gamma_{eg1}$ individually present the dephasing rates for $|e_{c,t}\rangle\to|0_{c,t}\rangle$ and $|e_{t}\rangle\to|1_{t}\rangle$  transitions.} First, by assuming $\gamma_{eg}=\gamma_{eg0}=\gamma_{eg1}$,
Fig. \ref{Fig6.flurabi}(b) shows the gate infidelity in the presence of different phase noises with the probe and coupling lasers. The value $\gamma_{re(eg)}$ can be computed by an independent measured single-atom Rabi frequency of 480 nm(780 nm) lasers and is given in the range of $2\pi\times [0,100]$ kHz. To account for the randomness of phase noise, we adopt stochastic values from $[0,\gamma_{re}]$ and $[0,\gamma_{eg}]$ for each realization and compute the average value after sufficient realizations. \textcolor{black}{Similarly, Fig. \ref{Fig6.flurabi}(c) presents the gate infidelity caused by the fluctuation in the relative phase between the two Raman lasers where the dephasing rates for $|e_{t}\rangle\to |0_{t}\rangle$ and $|e_{t}\rangle\to |1_{t}\rangle$ transitions are respectively denoted as $\gamma_{eg0}$ and $\gamma_{eg1}$(when $\gamma_{re}=0$).} From Fig. \ref{Fig6.flurabi}(b-c) we see the effect of laser phase noise is more influential than the intensity noise which 
determines the limitation on the observed gate fidelity in experiment \cite{PhysRevLett.123.230501}. When all dephasing rates $\gamma_{re}$, $\gamma_{eg0}$, $\gamma_{eg1}$ are stochastic in the range of $2\pi\times [0,100]$ kHz we find the gate infidelity reaches \textcolor{black}{$\mathcal{F}-\bar{\mathcal{F}}(\gamma_{re},\gamma_{eg0},\gamma_{eg1})\approx 7.79 \times 10^{-3}$} after running 500 simulations.

{\it Fluctuation of the two-photon detuning -} The two-photon detuning $\delta(t)$ also encounters an appropriate time-dependent modulation in our protocol which makes the adiabatic population transfer smoother. In reality, this modulation can be fluctuated by external magnetic fields or excitation lasers. Thus we give a random fluctuation $\delta\Delta/2\pi \in [0,100]$ kHz to $\delta(t)$ leading to
\begin{equation}
    \delta(t)\rightarrow \delta(t)+\delta\Delta.
\end{equation}
 The distribution of $\delta\Delta$ agrees with a uniform function in the range of $[-\delta\Delta,+\delta\Delta]$ in accordance with most experiments. 
 After 500 repeated simulations for each $\delta\Delta$ we show the average infidelity as a function of $\delta\Delta$ in Fig. \ref{Fig6.flurabi}(d). Luckily, the gate error caused by a fluctuated two-photon detuning has negligible effect $\sim 10^{-5}$ on the system.

\subsection{Influence from finite interaction strength}
\label{subsectionc}

\begin{figure}%FIGURE 
\includegraphics[width=0.47\textwidth]{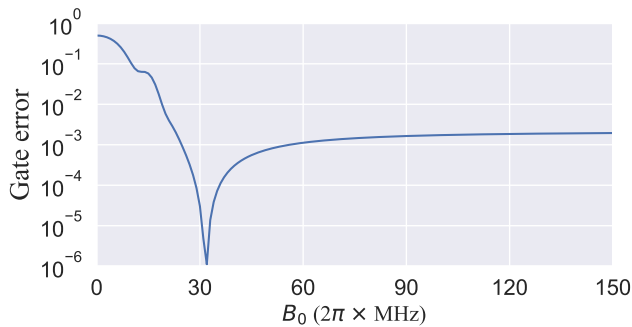} 
\caption{Gate error caused by the variation of the finite dipole-dipole exchange interaction $B_0$, and the valley value represents the case for $B_0/2\pi\approx32$ MHz at which numerical optimization is performed.}
\label{Fig6.flub}
\end{figure}

We also note that the interaction for  $|79D;79D\rangle\rightleftarrows|80P;78F\rangle$ is of anisotropic feature where $B_0 = C_3(\theta)/r_0^3$ and $C_3 \propto (1-3 {\rm cos}^2\theta)$ \textcolor{black}{\cite{PhysRevA.92.020701}}. $\theta$ is the angle between interaction axis $\vec{r}_0$ and quantization axis $\hat{x}$ and $r_0$ denotes the control-target distance. When $\theta=90^{\circ}$ and $r_0 = 9$ $\mu$m, $B_0/2\pi\approx32$ MHz as used in the paper. In reality $B_0$ depends on $\theta$ and $r_0$ so we tune $B_0/2\pi$ in the range of $[0,150]$ MHz in order to find out the infidelity of the gate with respect to a finite interaction strength $B_0$. As verified by Fig.\ref{Fig6.flub}, thanks to the parameter optimization the gate error reaches its minimum $\sim 1.1 \times 10^{-6}$ at $B_0/2\pi=32$ MHz. While further increasing $B_0$ by adjusting $\theta$ or $r_0$, this error keeps a steady value $\sim 2 \times 10^{-3}$ which is larger than the minimal value by three orders of magnitude. Oppositely, if one reduces $B_0$, the gate error quickly grows to an unacceptable level.
From this result, we stress that a strong blockade interaction is not necessary and our protocol is hence not a blockade gate. 
In other words, for a pair of relatively distant atoms with reduced interaction strengths, one still can obtain specific waveforms to maintain a reasonable gate fidelity. {\it E.g.} for $r_0$ = 17 $\mu$m corresponding to $B_0/2\pi \approx 5.0$ MHz the raw gate fidelity remains $\mathcal{F}\sim 0.995$ with optimal pulses(not shown).
This scheme advantage allows for the preservation of high-fidelity quantum gates on larger-scale atomic arrays.

\section{Discussion and Conclusion}

\begin{table}
\caption{\label{tab:table1} Error budget on the raw gate fidelity $\mathcal{F}=0.9987$ in case (i) of Table.\ref{tab:table3}. For representing a practical two-qubit CNOT gate
every error number is obtained through an average over repeated realizations
and values in parenthesis mean the maximal amplitudes of fluctuation
(see detailed discussions in Sec.V).}
\begin{ruledtabular}
\setlength{\tabcolsep}{0.5mm}{
\renewcommand{\arraystretch}{1.4}
\begin{tabular}{cc}
Quantity&Error Budget \\
\hline
\multicolumn{2}{c}{\it Caused by finite lifetime of energy levels(intrinsic)}\\
%\makecell{\it Caused by finite lifetime of energy levels(intrinsic)} & \\

\makecell{Spontaneous decay(Rydberg states)} &6.44 $\times$ $10^{-4}$\\

\makecell{Spontaneous decay(intermediate states)} &6.24 $\times$ $10^{-4}$\\
\hline
Finite blockaded interaction(intrinsic)\\($B_0/2\pi=32$ MHz) & 1.11 $\times$ $10^{-6}$\\
\hline
\multicolumn{2}{c}{\it Caused by finite atomic temperature at $T_a=30$ $\mu$K(technical)}\\
%{\it Caused by finite atomic temperature at $T_a=30$ $\mu$K(technical)} & \\
Fluctuated atomic position &3.81 $\times 10^{-6}$\\

Doppler dephasing error &1.61 $\times 10^{-5}$\\

Inhomogeneous Rabi frequency \\($\omega_r=7.8$ $\mu$m, $\omega_b=8.3$ $\mu$m) & 2.06 $\times 10^{-7}$\\
\hline
\multicolumn{2}{c}{\it Caused by excitation lasers(technical)}\\
%{\it Caused by excitation lasers(technical)} & \\
Laser intensity ($\delta\Omega_{r(b)}/2\pi=10$ MHz) &2.37 $\times 10^{-3}$\\

\textcolor{black}{Laser phase($\gamma_{eg0(1),re}/2\pi=50$ kHz)} &\textcolor{black}{4.22 $\times$ $10^{-3}$}\\

Two-photon detuning($\delta\Delta/2\pi=100$ kHz) &4.31 $\times$ $10^{-5}$ \\

\end{tabular}
}
\end{ruledtabular}
\end{table}

Before ending, we discuss the practical performance of two-qubit CNOT gates with experimental conditions. Table \ref{tab:table1} shows that the dominant gate errors arise from the excitation lasers, which leads to the variation of laser intensities and phases contributing to the gate error at the level of $\sim 10^{-3}$. So our protocol based on optimized pulses remains relatively sensitive to the fluctuation of excitation lasers. In addition, the atomic motion and dephasing caused by finite temperature($\sim30 \mu$K) leads to atomic position variations in the optical traps, contributes negligible $10^{-7}\sim10^{-5}$. Errors due to the finite radiative lifetime of Rydberg and intermediate states contribute the whole intrinsic error $(6.44+6.24)\times 10^{-4}\approx0.0013 =  1-\mathcal{F}$. After considering various technical imperfections in Table \ref{tab:table1}, we determine a more conservative lower bound on the predicted gate fidelity which is $\mathcal{F}>0.992$. The potential for a higher-fidelity Rydberg gate resorts to further technical improvements especially in laser(intensity and phase) noise reduction \textcolor{black}{\cite{Akerman_2015}}.

In conclusion, we have demonstrated a straightforward route to implement a two-qubit CNOT gate using a native two-photon Rydberg excitation. We explicitly show that, with smoothly-tuned Gaussian pulses obtained by numerical optimization, 
the intrinsic gate error can be merely determined by the spontaneous decays from the Rydberg and the intermediate states, leading to a raw gate fidelity as high as $0.9987$. By modeling various experimental error sources, we
have identified that the major obstacle due to technical imperfections arises from the excitation laser noises that make the optimal pulses fluctuated. 
Other technical limitations have been safely suppressed to a negligible level.
After a very conservative estimation by taking account of intrinsic and technical errors, the theoretically-predicted lower bound for fidelity is $0.992$ which is closer to the reported numbers by current laboratories.

Our proposal provides a promising realization of a more robust two-qubit Rydberg CNOT gate decreasing the gap between the theoretical fidelity prediction and the measured gate fidelity which contributes a practical proposal for Rydberg quantum gates. In addition, we show a faster and higher-fidelity gate ($T_g\sim 0.1$ $\mu$s, $\mathcal{F}\geq0.9929$) can be attainable via a stronger coupling to enhance the intermediate-Rydberg-state transition, 
which is primarily due to the reduction of laser phase noise within a shorter duration \textcolor{black}{\cite{Muller2011}}. More discussions about this high-speed gate proposal can be found in Appendix \ref{appendixb}. 
{\it \textcolor{black}{While preparing this manuscript we become aware of the significant advance recently by the Harvard group has realized two-qubit $C_Z$ gates with 0.995 fidelity closing the gap to other competitive quantum-computing platforms for the first time \cite{evered2023highfidelity}.}}

 \begin{acknowledgments}
We acknowledge financial support from the Innovation Program for Quantum Science and Technology 2021ZD0303200; the National Key Research and Development Program of China under Grant No. 2016YFA0302001;
by the NSFC under Grants No. 12174106, No.11474094,  No.11104076 and No.11654005, by the Science and Technology Commission of Shanghai Municipality under Grant No.18ZR1412800, by the Shanghai Municipal Science and Technology Major Project under Grant No. 2019SHZDZX01 and the Shanghai talent program.
\end{acknowledgments}

\appendix

\section{\textcolor{black}{Realistic Scheme Setup}}
\label{appendix}

\begin{figure}%FIGURE 
\includegraphics[width=0.47\textwidth]{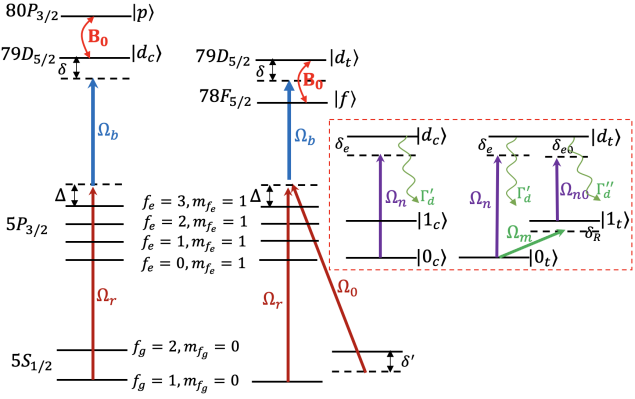} 
\caption{Atomic energy levels based on a realistic protocol with intermediate hyperfine states. Inset: A reduced model without intermediate states.}
\label{Fig7.hyperfine}
\end{figure}

\textcolor{black}{Taking note of multiple intermediate states for realistic case of Rydberg excitation, we extend the simple setup(Fig.\ref{Fig1.model}a) in the main text as follows. Due to the nuclear spin $I=\frac{3}{2}$ for $^{87}$Rb atoms the hyperfine energy levels of intermediate $|5P_{3/2}\rangle$ state are $|f_e,m_{f_e}\rangle=|0\sim3,1\rangle$ where the choice of $m_{f_e}$ depends on 
the dipole selection rules $\Delta m_f=+1$ with $\sigma^{+}$-polarized lasers between ground and intermediate states. The hyperfine splitting of Rydberg states can be ignored.}

\textcolor{black}{For this realistic protocol as in Fig.\ref{Fig7.hyperfine}, 
the full Hamiltonian taking into account all hyperfine intermediate states is explicitly described by 
\begin{eqnarray}
\mathcal{H}_{t} &=&
    \frac{1}{2}\sum_{f_e}[\Omega^{f_e}_r\left\vert 0_t\right\rangle\left\langle f_e\right\vert + \Omega^{f_e}_0 \left\vert 1_t\right\rangle\left\langle f_e\right\vert 
    + \Omega^{f_e}_b\left\vert f_e\right\rangle \left\langle d_t\right\vert] \nonumber\\
    &-&\sum_{f_e}\Delta_{f_e}\left\vert f_e\right\rangle \left\langle f_e\right\vert 
    -\delta^\prime\left\vert 1_t\right\rangle\left\langle 1_t\right\vert-\delta\left\vert d_t\right\rangle\left\langle d_t\right\vert +h.c.  \nonumber 
\end{eqnarray}
 for the target atom and is 
\begin{eqnarray}
    \mathcal{H}_{c} &=& \frac{1}{2} \sum_{f_e}[\Omega^{f_e}_r \left\vert 0_c\right\rangle \left\langle f_e\right\vert + \Omega^{f_e}_b \left\vert f_e\right\rangle \left\langle d_c\right\vert ] - \sum_{f_e}\Delta_{f_e}\left\vert f_e\right\rangle \left\langle f_e\right\vert\nonumber\\
     &-&\delta\left\vert d_c\right\rangle \left\langle d_c\right\vert+h.c.
     \nonumber
\end{eqnarray}
 for the control atom, where $\Delta_{f_e}=\Delta-E_{\mathrm{hfs}}(f_e)$ treats as the detuning of each hyperfine $|f_e,1\rangle$ state and $\Delta$ is the center-of-mass detuning from intermediate states. $\delta^\prime$ is an extra two-photon detuning in the Raman process, $\Omega^{f_e}_{r,0}$ stand for the Rabi frequencies between ground and hyperfine intermediate states, $\Omega^{f_e}_{b}$ is the Rabi frequency between hyperfine intermediate and Rydberg states. Note that the hyperfine level $|0,1\rangle$ is uncoupled to other states.}

\textcolor{black}{In order to accelerate the pulse optimization, we employ a reduced model by adiabatically eliminating hyperfine intermediate states for $|\Delta|\gg |\Omega^{f_e}_{r,b,0}|$ as presented in Fig.\ref{Fig7.hyperfine}(inset). The two-photon Rabi frequencies for the ground-Rydberg transitions $|0_{c,t}\rangle\to|d_{c,t}\rangle$ and $|1_{t}\rangle\to|d_{t}\rangle$ are respectively
\begin{equation}
 \Omega_{n} = \sum_{f_e}\frac{\Omega^{f_e}_r \Omega^{f_e}_b}{2\Delta_{f_e}}, \Omega_{n0} = \sum_{f_e}\frac{\Omega^{f_e}_0 \Omega^{f_e}_b}{2\Delta_{f_e}}  
\end{equation}
 and for the Raman transition $|0_{t}\rangle\to|1_{t}\rangle$ is 
\begin{equation}
\Omega_m=\sum_{f_e}\frac{\Omega^{f_e}_r\Omega^{f_e}_0}{2\Delta_{f_e}}   
\end{equation}
Correspondingly, the two-photon detunings take forms of
\begin{eqnarray}
\delta_{e}&=&\sum_{f_e}\frac{{(\Omega^{f_e}_r)}^2-{(\Omega^{f_e}_b)}^2}{4\Delta_{f_e}}+\delta \nonumber\\
\delta_{e0}&=&\sum_{f_e}\frac{{(\Omega^{f_e}_0)}^2-{(\Omega^{f_e}_b)}^2}{4\Delta_{f_e}}+\delta-
 \delta^{\prime} \nonumber\\
\delta_R&=&\sum_{f_e}\frac{{(\Omega^{f_e}_r)}^2-{(\Omega^{f_e}_0)}^2}{4\Delta_{f_e}}+\delta^{\prime}
\end{eqnarray}
including the influence of differential ac stark shifts $\sum_{f_e}(\Omega_i^{f_e})^2/4\Delta_{f_e}$($i=r,0,b$) for states $|0_{c,t}\rangle$, $|1_t\rangle$, $|d_{c,t}\rangle$ respectively. In the calculation the Rabi frequencies of the drives are given by 
\begin{equation}
\Omega^{f_e}_{r}=\alpha_r^{f_e}\Omega_r(t),\Omega^{f_e}_{0}=\alpha_0^{f_e}\Omega_r(t)
\end{equation}
where coefficients $\alpha^{f_e}_{r,0}$ are the dipole matrix elements [see Table.\ref{tab:tablex}]
and $\Omega^{f_e}_b \approx \Omega_b(t)$ for simplicity.} 

\begin{table}
\caption{\label{tab:tablex} Coefficients of dipole matrix elements and the corresponding detuning of each hyperfine intermediate state.}
\begin{ruledtabular}
\setlength{\tabcolsep}{0.5mm}{
\renewcommand{\arraystretch}{1.4}
\begin{tabular}{c|ccc}
$|f_e,m_{f_e}\rangle$&$|1,1\rangle$&$|2,1\rangle$&$|3,1\rangle$ \\
\hline
$\alpha^{f_e}_r$& $\sqrt{\frac{5}{24}}$ &$\sqrt{\frac{1}{8}}$ & -\\
$\alpha^{f_e}_0$&$\sqrt{\frac{1}{120}}$ & $\sqrt{\frac{1}{8}}$& $\sqrt{\frac{1}{5}}$\\
$\Delta_{f_e}/2\pi$(GHz) & 4.0 & 3.843 & 3.576 \\
\end{tabular}
}
\end{ruledtabular}
\end{table}

\textcolor{black}{To account for the gate performance, 
the spontaneous decays from intermediate states should be effectively incorporated into the Rydberg level via
\begin{eqnarray}
     \Gamma_d^\prime &=& \Gamma_d + \sum_{f_e}\frac{{(\alpha_r^{f_e}\Omega_r)}^2+\Omega_b^2}{4(\Delta_{f_e})^2}\Gamma_e \nonumber\\
     \Gamma_d^{\prime\prime} &=& \Gamma_d + \sum_{f_e}\frac{{(\alpha_0^{f_e}\Omega_r)}^2+{\Omega_b^2}}{4(\Delta_{f_e})^2}\Gamma_e 
\end{eqnarray}
and we apply native Rabi frequencies $\Omega_r(t)$, $\Omega_b(t)$ and detuning $\delta(t)$ smoothly vary as
\begin{eqnarray}
     \Omega_{r}(t) &=& \Omega_{r}^{\max}e^{-\frac{(t-T_g/2)^2}{2\omega_{r}^2}} \nonumber\\
     \Omega_{b}(t) &=& \Omega_{bm}e^{-\frac{(t-T_g/2)^2}{2w_b^2}}+K \nonumber\\
     \delta(t) &=& \delta_{0} + \delta_{1}\cos(\frac{2\pi t}{T_g}) + \delta_{2}\sin(\frac{\pi t}{T_g}) 
 \end{eqnarray}
while the Raman detuning $\delta^\prime$ is suitably chosen to maximally compensate the ac stark shifts.}

\begin{figure}
    \includegraphics[width=0.47\textwidth]{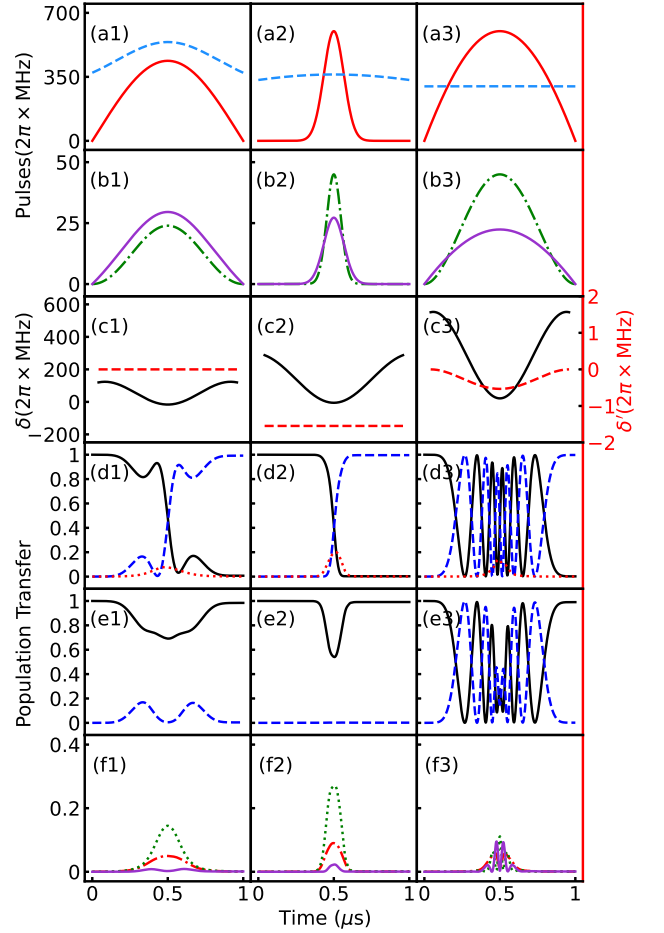} 
    \caption{\textcolor{black}{Optimal performance of a two-photon CNOT gate based on a realistic protocol with hyperfine intermediate states. From left to right, the two-photon detuning $\delta^\prime$ in the Raman transition is estimated to be zero, a fixed number $\delta^\prime/2\pi=-1.55$ MHz and a suitable time-dependent function $\delta^\prime(t)$ that can exactly compensate the differential ac stark shifts ($\delta_R\equiv0$). All linetypes and denotations are same as in Fig.\ref{Fig3.improvement} except for (c1-c3) $\delta$ and $\delta^\prime$ are denoted as black-solid and red-dashed lines, respectively.} }  
    \label{Fig8.appendix}
\end{figure}

\textcolor{black}{\begin{table*}
\caption{\label{tab:table4}{Coefficients of the optimized laser pulses $\Omega_r(t),\Omega_b(t)$ and the two-photon detuning $\delta(t)$ in Fig.\ref{Fig8.appendix} under different $\delta^{\prime}$ values. }}
\begin{ruledtabular}
\setlength{\tabcolsep}{0mm}{
\begin{tabular}{ccccccccccc}
\multicolumn{1}{c}{
}&\multicolumn{2}{c}{$\Omega_r(t)/2\pi$}&\multicolumn{3}{c}{$\Omega_b(t)/2\pi$} &\multicolumn{3}{c}{$\delta(t)/2\pi$}&$\mathcal{F}$\\
\hline
$\delta^\prime$(MHz)&$\Omega^{max}_r$(MHz)&$\omega_r$($\mu$s)&$\Omega_{bm}$(MHz)&$\omega_b$($\mu$s)&$K$(MHz)&$\delta_0$(MHz)&$\delta_1$(MHz)&$\delta_2$(MHz)&$\quad$\\
%\cmidrule(r){2-3}\cmidrule(r){4-6}\cmidrule(r){7-9}
\hline
$0$&438&0.4758&249.27&0.3341&291.14&24.41&95.49&54.65&0.9943\\
\hline
$-2\pi\times1.55$&600&0.0662&230.28&0.9427&132.83&192.27&95.19&-102.64&0.9969\\
\hline
$-\sum_{f_e}\frac{(\Omega_r^{fe})^2- (\Omega_0^{fe})^2}{4\Delta_{fe}} $ &600&0.7710&4.46&0.9807&294.11&251.70&300&69.88&0.9971\\
\end{tabular}
}
\end{ruledtabular}
\end{table*}}

\textcolor{black}{Numerical estimation adopts the master equation method in Appendix \ref{appendixa} with the reduced model, where the single-atom Hamiltonians $\mathcal{H}_{c}$, $\mathcal{H}_{t}$ are replaced by
\begin{eqnarray}
    \mathcal{H}_c &=& -\delta_e|d_c\rangle\langle d_c|+\frac{\Omega_n}{2}|0_c\rangle\langle d_c| +h.c.\nonumber\\
    \mathcal{H}_t &=& - \delta_{e0} |d_t\rangle \langle d_t| - \delta_R |1_t\rangle \langle 1_t| - \delta_e |d_t\rangle \langle d_t|+\frac{\Omega_n}{2}|0_t\rangle \langle d_t| \nonumber\\
     &+& \frac{\Omega_{n0}}{2} |1_t\rangle \langle d_t| + \frac{\Omega_m}{2} |0_t\rangle \langle 1_t|+ h.c. 
\end{eqnarray}
and the jump operators $L_{s1}$, $L_{s2}$ are 
\begin{eqnarray}
L_{s1}&=&\sqrt{\Gamma^{\prime}_d}(|0_c\rangle\langle d_c| +|1_c\rangle\langle d_c|) + \sqrt{\Gamma_p}(|0_c\rangle \langle p| + |1_c\rangle \langle p|) \nonumber\\
    L_{s2}&=&\sqrt{\Gamma^{\prime}_d}|0_t\rangle\langle d_t| + \sqrt{\Gamma^{\prime\prime}_d}|1_t\rangle\langle d_t| + \sqrt{\Gamma_f}(|0_t\rangle \langle f| + |1_t\rangle \langle f|) \nonumber
\end{eqnarray}}
% \begin{eqnarray}
%     L_{se1}&=&\sqrt{\Gamma^{\prime}_d}|0_c\rangle\langle d_c| + \sqrt{\Gamma^{\prime}_d}|1_c\rangle\langle d_c| + \sqrt{\Gamma_p}(|0_c\rangle \langle p| \nonumber\\
%     &+& |1_c\rangle \langle p|) \nonumber\\
%     L_{se2}&=&\sqrt{\Gamma^{\prime}_d}|0_t\rangle\langle d_t| + \sqrt{\Gamma^{\prime\prime}_d}|1_t\rangle\langle d_t| + \sqrt{\Gamma_f}(|0_t\rangle \langle f| \nonumber\\
%     &+& |1_t\rangle \langle f|)
% \end{eqnarray}

\textcolor{black}{In Fig. \ref{Fig8.appendix} we explicitly present the optimized gate performance under different choices of detuning $\delta^{\prime}$. Relevant parameters are given in Table.\ref{tab:table4}. When $\delta^{\prime}=0$ the differential ac stark shifts on the two-photon Raman transition is not compensated, we achieve a relatively lower fidelity $\mathcal{F}=0.9943$. Then we proceed to use a fixed Raman detuning to partially compensate the shifts yielding $\mathcal{F}=0.9969$ at $\delta^\prime/2\pi=-1.55$ MHz(optimal). As compared to the case $\delta^{\prime}=0$, data in Fig.\ref{Fig8.appendix}(d2-f2) show that the time-dependent state transfer becomes smoother with optimized adiabatic pulses where the intermediate-state population is maximally suppressed. In addition, taking note of the ac stark shifts that evolve dynamically during the pulse sequence, we apply a perfect compensation using a time-dependent $\delta^\prime(t)$ by letting $\delta_R\equiv0$. In this case, we can extract a competitive fidelity $\mathcal{F}=0.9971$ for the vanishing of ac shifts on the Raman transition. However, the third case suffers from significantly strong oscillations in the state transfer because of the two-photon resonance between $|0_t\rangle$ and $|1_t\rangle$.}

\textcolor{black}{Results above also imply that a fixed detuning to the Raman transition is sufficient to overcome the ac stark shifts yielding a high-fidelity state transfer \cite{PhysRevLett.129.200501}. In the main text, we present a simplified model without multiple intermediate states and focus on the ways of global pulse optimization. We have assumed that two Raman lasers have the same Rabi frequency 
$\Omega_r(t)=\Omega_0(t)$ as done by \cite{PhysRevLett.102.170502}, so that the differential ac stark shifts in the Raman process vanish, leading to $\delta^\prime=0$. Therefore, as shown in Fig.\ref{Fig1.model}a, two Raman pulses $\Omega_r(t)$ couple the ground states $|0_t\rangle$ and $|1_t\rangle$ with a zero two-photon detuning.}

\section{Numerical Method}
\label{appendixa}

As stated in the main text, we consider the gate implementation by trapping two $^{87}$Rb atoms with simplified atomic states $\{|0_{c(t)}\rangle$, $|1_{c(t)}\rangle$, $|e_{c(t)}\rangle$, $|d_{c(t)}\rangle$, $|p\rangle$, $|f\rangle\}$=$\{|5S_{1/2},F=1,m_F=0\rangle$, $|5S_{1/2},F=2,m_F=0\rangle$, $|5P_{3/2}\rangle$, $|79D\rangle$, $|80P\rangle$, $|78F\rangle\}$ in Fig.\ref{Fig1.model}. For $T_a\approx 30$ $\mu$K
the lifetime of $|e_{c,t}\rangle$ is 0.053 $\mu$s and of $|d_{c,t},p,f\rangle$ is 53.05 $\mu$s, giving rise to $\Gamma_e/2\pi\approx3.0$ MHz, $\Gamma_{d,p,f}/2\pi=3.0$ kHz. 
$B_0=C_3/r_0^3\approx 32$ MHz stands for the non-fluctuated dipole-dipole exchange interaction for $|d_cd_t\rangle\rightleftarrows|pf\rangle$ transition. To represent a native two-photon excitation, the transitions of $|0_c\rangle\to|e_c\rangle$ and $|0_t(1_t)\rangle\to|e_t\rangle$ are detuned from $|e_c\rangle$ or $|e_t\rangle$
by $\Delta/2\pi = 4.0$ GHz which is appropriately chosen to show the influence of intermediate-state decays. Note that we employ a straightforward method to optimize all time-dependent pulses (including laser waveforms $\Omega_{r(b)}(t)$ and two-photon detuning $\delta(t)$) at the expense of a time-consuming computing which can be readily for experimental demonstration.

Numerical solving of accurate population dynamics for each input computational state depends on the master equation method in which the effects of atom-light couplings and spontaneous decays are both included \textcolor{black}{\cite{PhysRevA.95.012708}},
\begin{equation}
    \frac{d{\rho}}{dt}=-i[\mathcal{H},{\rho}]+\mathcal{L}_s[{\rho}]+\mathcal{L}_d[{\rho}]
    \label{raa}
\end{equation}
with $\rho(t)$ the time-dependent density matrix and $\mathcal{H}$ the total Hamiltonian 
\begin{equation}
    \mathcal{H} = \mathcal{H}_c\otimes \mathcal{I} + \mathcal{I}\otimes \mathcal{H}_t + \mathcal{H}_{dd}.
\end{equation}
Here the denotations of $\mathcal{H}_c$, $\mathcal{H}_t$, $\mathcal{H}_{dd}$ can be found in Sec. II. Dissipator $\mathcal{L}_s[{\rho}]$ 
describing the effects of spontaneous emissions from Rydberg(denoted by rates $\Gamma_{d,f,p}$) and from intermediate(denoted by rate $\Gamma_{e}$) states, is given by
\begin{equation}
    \mathcal{L}_s[{\rho}] =  \sum_{j=1}^{2}[L_{sj}{\rho}L_{sj}^{\dagger}-\frac{1}{2}(L_{sj}^{\dagger}L_{sj}{\rho}+{\rho} L_{sj}^{\dagger}L_{sj})] 
\end{equation}
with jump operators
\begin{eqnarray} 
L_{s1}&=&\sqrt{\Gamma_{d}}(|0_{c}\rangle\langle d_{c}|+|1_{c}\rangle\langle d_{c}|)+\sqrt{\Gamma_{p}}(|0_{c}\rangle\langle p|+|1_{c}\rangle\langle p|) \nonumber\\
&+&\sqrt{\Gamma_{e}}(|0_{c}\rangle\langle e_{c}|+|1_{c}\rangle\langle e_{c}|) \nonumber\\
L_{s2}&=&\sqrt{\Gamma_{d}}(|0_{t}\rangle\langle d_{t}|+|1_{t}\rangle\langle d_{t}|)+\sqrt{\Gamma_{f}}(|0_{t}\rangle\langle f|+|1_{t}\rangle\langle f|) \nonumber\\
&+&\sqrt{\Gamma_{e}}(|0_{t}\rangle\langle e_{t}|+|1_{t}\rangle\langle e_{t}|)
\end{eqnarray} 
The dissipator $\mathcal{L}_{d}[\rho]$ (see Eq.(\ref{Ld}))
means the decoherence due to the limited laser linewidths, which is only used to reflect the average effect of laser phase noises.

 Starting from arbitrary computational state, we solve the population evolution which is governed by equation (\ref{raa}). The two-qubit gate fidelity is calculated by using
\begin{equation}
   \mathcal{F}=\frac{1}{4}\text{Tr}[\sqrt{\sqrt{\rho_{et}}{\rho}(t=T_g)\sqrt{\rho_{et}}}] 
\end{equation} 
 where $\rho_{et}$ is an etalon CNOT gate matrix
 \begin{equation} %开始数学环境
	\setlength{\arraycolsep}{5pt}
	\rho_{et}=\left(                 %左括号
	\begin{array}{cccc}   %该矩阵一共3列，每一列都居中放置
		1 & 0 & 0 & 0 \\  %第一行元素
		0 & 1 & 0 & 0 \\  %第二行元素
		0 & 0 & 0 & 1 \\  
		0 & 0 & 1 & 0 \\     
	\end{array}
	\right)   
	\label{rhoet}%右括号
\end{equation}
according to the inputs $\{|0_c0_t\rangle,|0_c1_t\rangle,|1_c0_t\rangle,|1_c1_t\rangle\}$ respectively and ${\rho}(t=T_g)$ is the realistic output matrix obtained at the end of actual-pulse implementation.

\section{Fast two-qubit CNOT gate }
\label{appendixb}

\begin{table}
\caption{\label{tab:table2}Error budget for case (iii) in Table.\ref{tab:table3}.}
\begin{ruledtabular}
\setlength{\tabcolsep}{2mm}{
\renewcommand{\arraystretch}{1.4}
\begin{tabular}{cc}
Quantity&Error Budget\\
\hline
\multicolumn{2}{c}{\it Caused by finite lifetime of energy levels(intrinsic)}\\

\makecell{Spontaneous decay(Rydberg states)} & 3.64 $\times$ $10^{-4}$\\

\makecell{Spontaneous decay(intermediate states)} & 1.17 $\times$ $10^{-3}$\\

\hline
Finite blockaded interaction\\($B_0/2\pi=32$ MHz) & 8.38 $\times$ $10^{-5}$\\

\hline
\multicolumn{2}{c}{\it Caused by finite atomic temperature at $T_a=30$ $\mu$K(technical)}\\

Fluctuated atomic position & 4.51 $\times 10^{-4}$\\

Doppler dephasing error & 3.17 $\times 10^{-6}$\\

Inhomogeneous Rabi frequency \\($\omega_r=7.8$ $\mu$m, $\omega_b=8.3$ $\mu$m) & 6.56 $\times 10^{-7}$\\
\hline
\multicolumn{2}{c}{\it Caused by excitation lasers(technical)}\\

Laser intensity ($\delta\Omega_{r(b)}/2\pi=10$ MHz) & 1.23 $\times 10^{-3}$\\

Laser phase($\gamma_{eg0(1),re}/2\pi=50$ kHz) & 2.12 $\times$ $10^{-3}$\\

Two-photon detuning($\delta\Delta/2\pi=100$ kHz) & 6.98 $\times$ $10^{-6}$ \\
\end{tabular}
}
\end{ruledtabular}
\end{table}

In Sec. III we show that our protocol has the ability to implement a high-speed CNOT gate within 0.1 $\mu$s, which is at the expense of a stronger coupling laser $\Omega_b(t)$ with its peak $2\pi\times 975$ MHz and width 0.7336 $\mu$s.
A stronger coupling laser can enhance the gate speed by reducing the time spent on the intermediate lossy levels. Remember the current mainstream blueprint for a CNOT gate requires several single-qubit rotations to convert a typical two-qubit $C_z$ gate into a more useful CNOT gate \textcolor{black}{\cite{PhysRevA.82.030306}}. This usually needs a longer duration. 
The potential for a faster and higher-fidelity CNOT gate makes neutral-atom computing platform more competitive with other trapped ions and superconducting qubits systems. 

To fully understand the performance of such a CNOT gate with its raw fidelity as high as $\mathcal{F}=0.9984$, we also account for the error estimate as shown in Table.\ref{tab:table2}.
As compared with the $1.0$-$\mu$s case(Table.\ref{tab:table1}), the raw fidelity is lowered by 0.0003(0.9987$\to$0.9984) which is mainly caused by the population remaining in the lossy intermediate states. Another important feature is the obvious reduction in laser phase noise that has been translated into one source of dephasing(except Doppler dephasing). A scheme using a shorter gate duration accompanied by a higher laser power can significantly decrease the laser phase noise from $4.22\times10^{-3}$ to $2.12\times10^{-3}$, eventually leading to a higher lower bound for the gate fidelity
$\mathcal{F}\geq 0.9929$.
This result, if cooperating with technology to suppress laser intensity and phase noises \textcolor{black}{\cite{Nazarova:08}}, may offer a feasible route to the experimental demonstration of ultrafast quantum computing in neutral-atom platforms.



\end{document}